\documentclass[english,reprint,prb]{revtex4-1}
\pdfoutput=1
\usepackage[T1]{fontenc}
\usepackage[latin9]{inputenc}
\usepackage{bm}
\usepackage{amsmath}
\usepackage{graphicx}
\usepackage{amssymb}
\newcommand{\beq}{\begin{equation}}
\newcommand{\eeq}{\end{equation}}
\newcommand{\beqa}{\begin{eqnarray}}
\newcommand{\eeqa}{\end{eqnarray}}

\makeatletter

\@ifundefined{textcolor}{}
{%
 \definecolor{BLACK}{gray}{0}
 \definecolor{WHITE}{gray}{1}
 \definecolor{RED}{rgb}{1,0,0}
 \definecolor{GREEN}{rgb}{0,1,0}
 \definecolor{BLUE}{rgb}{0,0,1}
 \definecolor{CYAN}{cmyk}{1,0,0,0}
 \definecolor{MAGENTA}{cmyk}{0,1,0,0}
 \definecolor{YELLOW}{cmyk}{0,0,1,0}
 }

\makeatother

\usepackage{babel}

\begin{document}

\bibliographystyle{apsrev4-1}

\title{Anderson localization and topological transition in Chern insulators}
\author{ Eduardo V. Castro$^{1, 2}$}
\author{M. Pilar L\'opez-Sancho$^{3}$}
\author{Mar\'{\i}a A. H. Vozmediano$^{3}$}

\affiliation{$^{1}$CeFEMA, Instituto Superior T\'{e}cnico, Universidade de 
Lisboa, Av. Rovisco Pais, 1049-001 Lisboa, Portugal}

\affiliation{$^{2}$Beijing Computational Science Research Center, Beijing 100084, China
}
\affiliation{$^{3}$Instituto de Ciencia de Materiales de Madrid, CSIC,
Sor Juana In\'es de la Cruz 3,  Cantoblanco,
E-28049 Madrid, Spain}
\date{\today}

\begin{abstract}
We analyze the topological transition and localization evolution of 
disordered two dimensional systems
with non trivial topology based on bipartite lattices. 
Chern insulators with broken time reversal symmetry
show non standard behavior for disorder realizations selectively
distributed on 
only one of the sublattices. 
The Chern number survives to a much stronger disorder strength (one 
order of magnitude higher) than in the equally distributed disordered case 
and the final state in the strongly disordered case is metallic. 
\end{abstract}
\maketitle
\section{Introduction-summary}
\label{sec_intro}
%{\it Introduction-summary.}
The metal insulator transition and the associated issue of the existence or not of metals in two dimensions \cite{Anderson58} have fascinated physicists in the last century and have been the focus of intense debates 
and great advances in the understanding of low dimensional material physics. The interest of these materials is of special relevance today  due to the unusual applications expected following the experimental capability of synthesizing and manipulating layered compounds. \cite{GG13} But also the fundamental physical aspects are 
still providing surprises mostly around the Dirac physics based materials and their related topologically non-trivial features. The issue of how electrons in a metal become localized by disorder in the presence of non--trivial topology started  with the analysis of Landau levels in the integer quantum Hall effect and was rapidly adapted to other time reversal preserving topological matter. \cite{BHZ06,OFRM07,HK10,QZ11} %(a very neat description can be found in the introduction of \cite{OFRM07}). 
%By now it is assumed that the localization behavior of materials are determined by symmetry, topology, and  space dimensions. 
The standard classification of disordered classes \cite{AALR79,ATAF80} based on the Altland-Zirnbauer sets of random matrices  \cite{AZ97} was completed to include topological features and the ``tenfold way'' was set in Refs. \onlinecite{SRFL08,EM08}.

%Despite its long history, the localization properties of two dimensional systems remains being a fundamental topic in condensed matter physics. After the seminal work of P. Anderson \cite{Anderson58} is was understood that in a non-interacting two dimensional electron system at zero temperature in spacial dimension $D\geq 2$ and in the thermodynamic limit, the electronic wave function will be localized by disorder. In more realistic situations the scaling theory of localization allowed a classification of the localization behavior of materials into universality classes set by symmetry and space dimensionality  \cite{AALR79,ATAF80} based on the Altland-Zirnbauer sets of random matrices  \cite{AZ97}. The advent of topological insulators \cite{BHZ06,HK10,QZ11} provided a new class of delocalized states, the edge states, robust under disorder provided some discrete symmetries were preserved. The universality classes were then adapted to include the topological features and a ``tenfold way" classification  was set \cite{SRFL08,EM08}. 

Two dimensional (2D) topological insulators belong to class A (unitary) with time reversal symmetry broken or AII (symplectic) time reversal invariant. Prominent examples in these 
classes are the Haldane \cite{H88} and the Kane-Mele model. \cite{KM05} They are characterized by a quantized Chern (or spin Chern) number $C$. The localization transition in these cases is accompanied by a topological transition between a topological (band) insulator and a trivial (Anderson) insulator. The topological transition of these classes is assumed to be well understood. The standard mechanism is referred to as ``levitation and annihilation'' of extanded states.\cite{OAN07} For moderate disorder, an impurity band of localized state forms inside the gap and simultaneously the high energy states in the edges of the conduction and valence bands  start to localize. As disorder increases, the gap is totally populated by localized states and the bulk extended states above and below the Fermi level carrying the Chern number shift toward one another and annihilate
leading to the topological phase transition. The transition from a Chern to an Anderson insulator under a random  potential disorder has been analyzed in Ref. \onlinecite{PHB10}. In the standard classification there are also the chiral classes characterized by having sublattice symmetry. They are typically analyzed by adding random disorder to a bipartite lattice. \cite{GW91,KOPM12} Graphene belongs to this class and its localization properties have been studied at length. \cite{OTetal10,HSetal14,FUH14,OPK14} The localization behavior of Chiral classes is more complicated  but they do not have associated topological invariants in two dimensions.

In this work  we focus on the sublattice structure of the honeycomb 
lattice underlying 2D topological insulators and show 
that the critical disorder which 
signals the appearance of the final regime has an extraordinary sensitivity 
to the disorder selectivity over sublattices/orbitals. 
The Anderson transition in the chiral classes is particularly interesting with interactions playing a role \cite{KOPM12} but, as mentioned, there are no topological compounds in this class in 2D. 
We see that, for disorder realizations affecting only one of the sublattices, 
the Chern number survives to a much stronger disorder strength (one 
order of magnitude higher) than in the equally disordered case and the localization transition 
never takes place. We exemplify our findings with the Haldane model \cite{H88}
with two different types of disorder: Anderson disorder and vacancies. It is interesting
to note that, although there are qualitative differences on the spectral evolution as disorder
is increased, the two cases behave similarly and the important difference lies in the two possible disorder realizations.

%We start with a topological band insulator with a well defined gap between valence and conduction bands that have opposite Chern numbers. Disorder induces localised states in the gap that eventually evolve to form an impurity band with zero Chern number. Simultaneously the states in the edges of the valence and conduction bands become localised. The Chern number is carried by the bulk extended states remaining in the valence and conduction bands among a set of disorder--localized states. For a critical value of disorder a Òlevitation and annihilationÓ process occur where the bulk extended states above and below the Fermi level levitate toward one another in the energy spectrum and then annihilate upon collision, leading to topological phase transitions.

We visualize the topological transition as a transition from the honeycomb 
(topologically non--trivial for finite Haldane mass) to the triangular (trivial) lattice within 
the same symmetry class. The ``localization transition'' is more 
complicated but in the limiting cases it occurs between a band 
insulator (topologically non trivial) and a metal, again without 
changing symmetry or dimensionality.

\section{Model and methods}
%{\it Model and methods.}
We use the Haldane model \cite{H88} as a generic example of a topologically
non trivial system in symmetry class A. It was initially proposed as a tight binding
model in the honeycomb lattice with complex next nearest neighbor (NNN) hoppings carefully adjusted
so as to mimic a magnetic flux flowing in opposite directions through the two triangular 
sublattices. It was build to show the possibility to have Hall conductivity in zero external
magnetic field and it is nowadays the prototype of a modern topological insulator. 
Interestingly, a physical realization of the model has been implemented in 
Ref.~\onlinecite{JMetal14} (see also Ref.~\onlinecite{W13}).
The Haldane model tight binding Hamiltonian can be written as 
%cast in the
%form (see Fig.~\ref{fig:lattice}),
\beqa
H  &=  -t\displaystyle\sum_{\left\langle i,j\right\rangle }c_i^{\dagger}c_j+%\\ \nonumber
-t_2\sum_{\left\langle \left\langle i,j\right\rangle \right\rangle }e^{-i\phi_{ij}}c_i^{\dagger}c_j\\ \nonumber
&+  M\displaystyle\sum_i \eta_i c_i^{\dagger}c_i +\mbox{H.c.},
\eeqa
where $c_i=a,b$ is defined in the two triangular sublattices $A,B$ that form the honeycomb lattice. 
The first term $t$ represents a standard
real nearest neighbor hopping that links the two triangular sublattices.
The next term represents a complex NNN hopping $t_2 e^{-i\phi_{ij}}$
acting within each triangular sublattice with a 
phase  $\phi_{ij}$ that has
opposite signs $\phi_{ij}=\pm\phi$.
This term breaks time--reversal symmetry and opens a non--trivial topological gap
at the Dirac points. We have done our calculations for the simplest case $\phi=\pi/2$.
The last term represents a staggered potential($\eta_i=\pm 1$). It breaks 
inversion symmetry and opens a trivial gap at the Dirac points. 
%The details of the tight binding Hamiltonian are described in the supplementary information \ref{sec_model}.  

We will discuss two types of disorder: Anderson and vacancy disorder equally or selectively
distributed among the two sublattices. Potential (Anderson) disorder is implemented by adding to the Hamiltonian the term $\sum_{i\in A,B}\varepsilon_{i}c_{i}^{\dagger}c_{i}$,
with a uniform distribution of random local energies, $\varepsilon_{i}\in[-W/2,W/2]$. For selective disorder the sum runs only over one sublattice. The same is done for vacancies which are introduced by removing a given lattice site and the  hopping to its neighbors.

The Haldane model belongs to symmetry class~A where the different topological
phases can be characterized by a $\mathbb{Z}$--topological number,
the Chern number $C$.\cite{SRFL08} 
In the clean system it can be computed from the single particle Bloch states $u_n({\bf k})$ as:
\beq
C_n=\frac{1}{2\pi}\int_S\Omega_z^n({\bf k}) dS,
\eeq
where the integral is over the unit cell $S$ and  $\Omega_z^n({\bf k})$ is the $z$
component of the Berry curvature,
$ \Omega^n(\mathbf{k})= \nabla_{\bf k}\wedge {\cal A}_n(\mathbf{k})$  defined from the
Berry connection:
${\cal A}_n({\bf k})=\left<u_n({\bf k})\vert -i\nabla_{{\bf k}}\vert u_n({\bf k})\right>$.
When translational invariance is broken by disorder or any other perturbation, the Chern number
has to be computed numerically in real space. 
The problem has been addressed in the literature and 
a number very efficient numerical techniques 
based on the use of twisted boundary conditions are now
available. \cite{PHB10,ZYetal13}
From the computational point of view an efficient approach was developed by Fukui et al.\cite{FHS05}
where Chern numbers are calculated on the basis of a discretized Brillouin zone.
Based in this technique  Zhang el al. \cite{ZYetal13} have recently
proposed  an efficient method  applying  a
coupling-matrix approach to calculate the Chern number in finite disordered systems.
The method  is faster than previous ones allowing to obtain the Chern number
with a single diagonalization for the $N_1 \times N_2$ system, if $N_1$, $N_2$ are large enough.
The technique has been probed to give very good results for the Haldane model.
In the present work we have calculated the Chern number for the bulk spectrum following
the coupling-matrix approach described in Ref.~\onlinecite{ZYetal13}.

The bulk density of states (DOS) have been computed using the recursion Green's function
method as used for example in Refs.~\onlinecite{CLV10,CdS08}.

\section{Results}
\label{sec_results}
%{\it Results.}

\begin{figure}
\begin{centering}
\includegraphics[width=0.98\columnwidth]{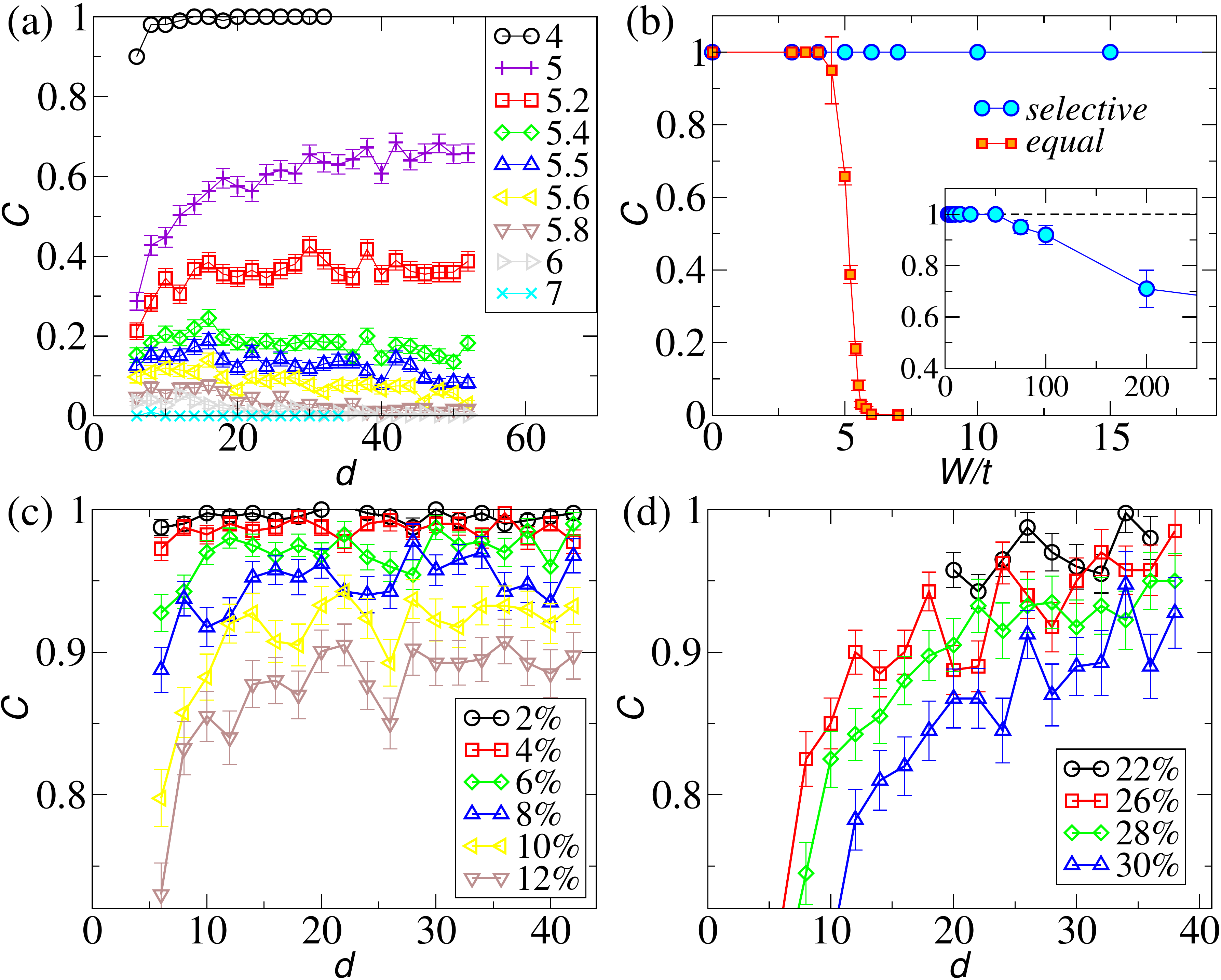}
\par\end{centering}
\caption{\label{Chern} Evolution of the Chern number for Anderson disorder [(a), (b)] and 
vacancy disorder [(c), (d)]. 
(a) Chern number as a function of the linear
lattice size $d$ for disorder realizations equally distributed
among the two sublattices. The different curves correspond to different disorder strengths $W$ in units of $t$. The error bars 
stand for the standard deviation due to disorder average.
(b) A comparison of the behavior of the Chern number as a function
of the disorder strength for the two types of disorder: 
equally distributed (red squares) and selective disorder (blue circles). The inset shows the
-- much higher -- values of disorder strength for which the Chern number ceases to be quantized.
(c) Chern number for equally diluted sublattices. (d) Selected dilution.}
\end{figure}

\begin{figure}
\begin{centering}
\includegraphics[width=0.9\columnwidth]{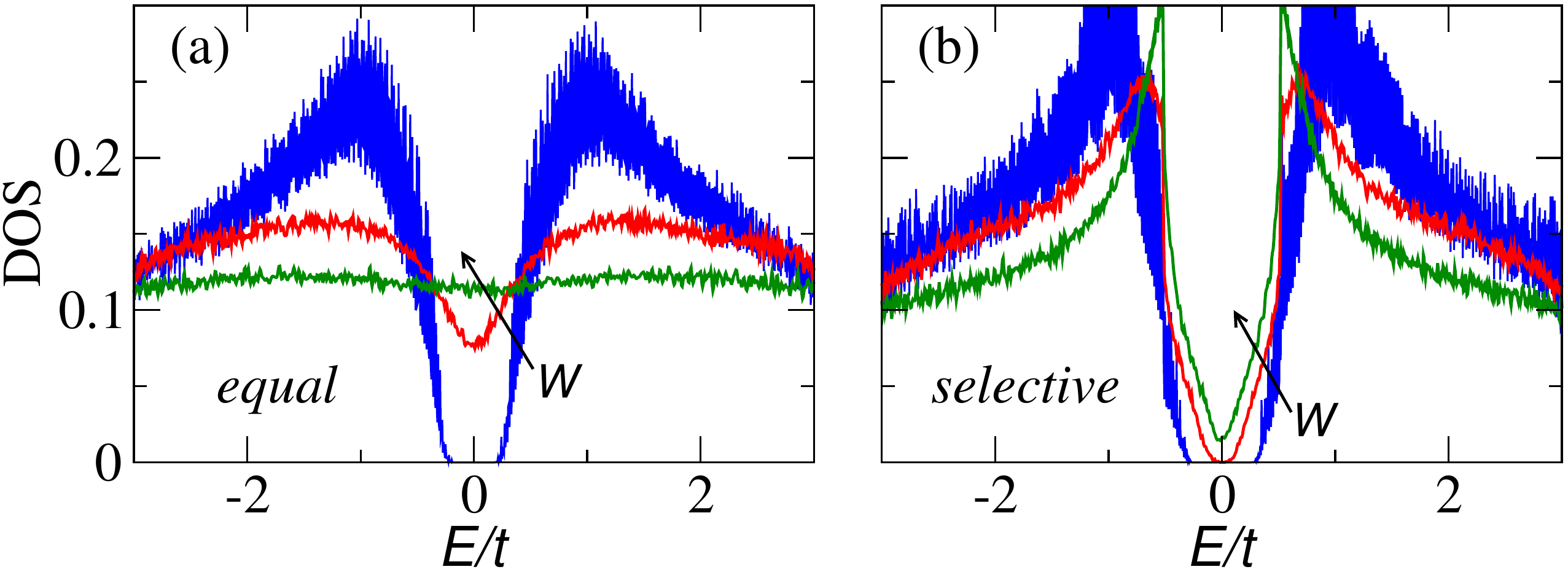}
\par\end{centering}
\caption{\label{DOSand} DOS of the Haldane model with Anderson disorder with equal (a)
and selective (b) disorder distribution. The three curves refer to different values of the 
disorder strength: $W/t=2, 4, 6$. The arrows indicate the effect of increasing disorder.}
\end{figure}

\begin{figure}
\begin{centering}
\includegraphics[width=0.98\columnwidth]{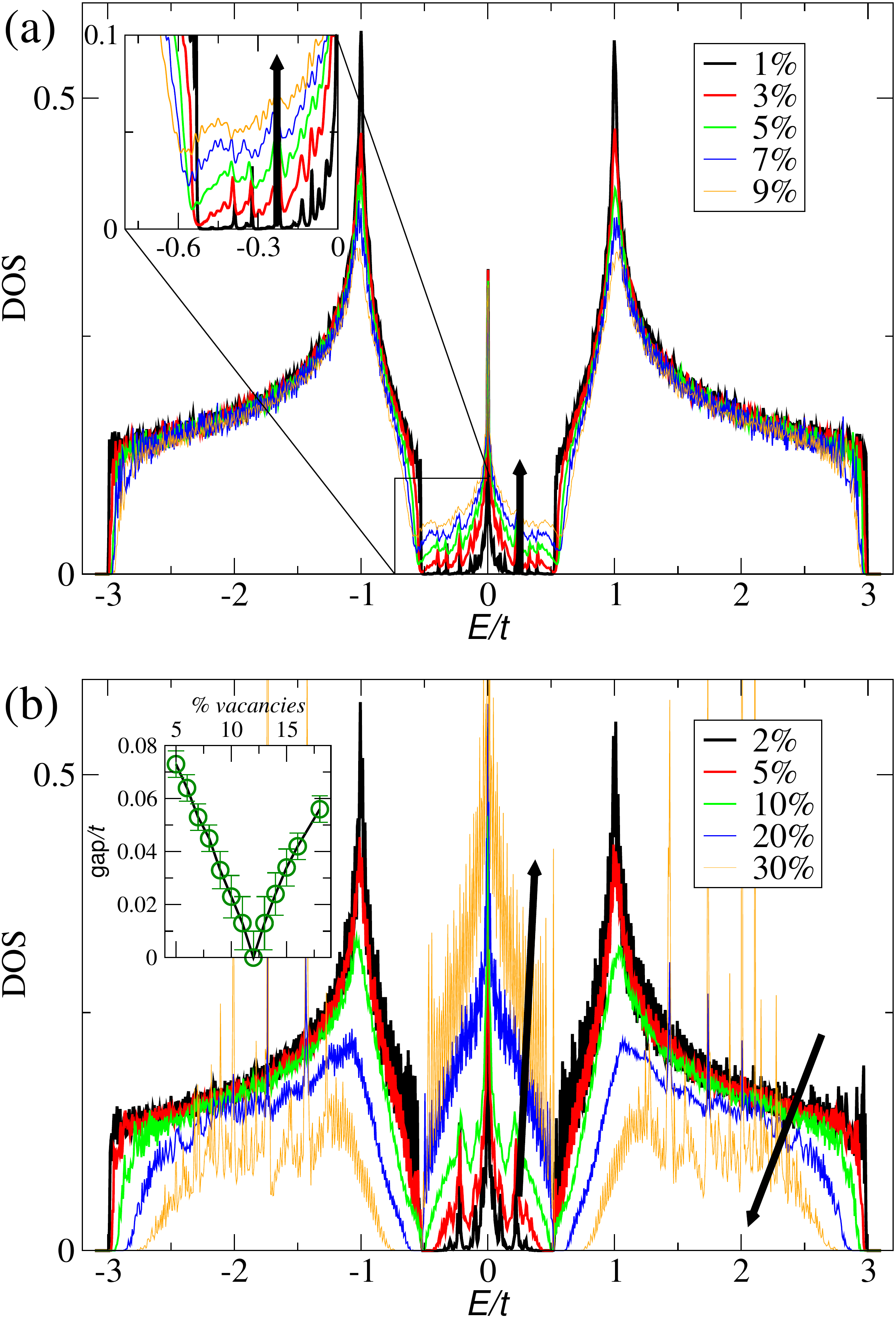}
\par\end{centering}
\caption{\label{DOSvac} DOS of the Haldane model for
vacancies equally distributed among the
two sublattices (a) and on one sublattice only (b). 
The insets show the behavior of the gap as a function of the density of vacancies for the two situations. For equally distributed vacancies the gap closes for a critical density 
$nv \sim 3 \% $, as can be seen in the zoomed in region. For selected dilution, the gap closes at the critical density $nv \sim 12\% $ but reopens again for $nv > 12\% $. Arrows indicate the effect of increasing disorder.}
\end{figure}

As shown by many authors, a single vacancy induces a midgap bound state
in the Haldane or Kane Mele model, \cite{dugaev2011impurity,SLetal11,gonzlez-iisathotqshp2012,NgImpurity,he2013zero,leeBandImp}
while a density of vacancies induces an impurity band inside the gap.\cite{leeBandImp} 
Increasing the density of vacancies enhances the
spectral weight inside the gap, which eventually closes for high enough
vacancy density. In contrast, potential disorder localizes states near the band edges
and generate a mobility edge. \cite{LR85} 
Lifshitz tails eventually overlap with incresing disorder and the gap closes.  
We see that Anderson disorder (local random potential) and vacancies show quantitative 
differences only. The crucial difference is found between the two disorder
realizations analyzed: selective and equally distributed in
the two sublattices. We summarize our results in Figs.~\ref{Chern}--\ref{DOSvac}.
The evolution of the Chern number with disorder strength encodes the topological transition.
The various cases are shown in Fig. \ref{Chern}.  The bulk DOS contains
information on the 
%localization behavior 
spectral gap
and is displayed in
Figs. \ref{DOSand} and  \ref{DOSvac}. 

We will first discuss the situation when impurities or vacancies 
are equally distributed in the two sublattices. 
In Fig.~\ref{Chern}(a) we show the behavior of the Chern number 
with potential disorder as a function
of the linear lattice size $d$ for increasing disorder strengths $W$ in units of $t$. 
%In the left hand side the disorder is equally distributed among the two sublattices. 
We see that the Chern number
is already well defined and stable at very small lattice sizes $d$. 
As shown in Fig.~\ref{Chern}(b), where we plot the Chern number\footnote {For convinience of language
we use the term Chern number even if the averaged result is not an integer. }
for the largest simulated lattice size $d$
 as a function of disorder strength $W$, the obained number ceases to be quantized 
at a mild disorder strength ($4 t<W_c<5 t$), a behavior  already found in the 
literature. \cite{SSWH05,ZYetal13} 
%The lower, left hand side  panel of Fig. \ref{Chern} shown the 
A  comparison of the DOS of the Haldane model with Anderson disorder 
is shown in Fig.~\ref{DOSand}(a) for  equal disorder and
in Fig.~\ref{DOSand}(b) for  selective disorder. The three curves shown in 
each panel refer to three disorder strengths $W/t=2,4,6$.
For weak disorder strength $W=2t$ (blue), well below the critical value
for the topological transition of the case of equally distributed disorder,
the two images are very similar. We see a well defined gap 
at zero energy and two peaks of similar intensity in the conduction and valence bands,
a situation compatible with the standard picture of the mobility edge. 
For a disorder strength $W=4t$ (red), close to the critical value for the topological transition  of the case of 
equally distributed disorder, the DOS shows a clear difference between the two disorder distributions. 
For the selectively distributed case in Fig.~\ref{DOSand}(b) the structure of the DOS is similar to the weak disorder case
while in the equally distributed case in Fig.~\ref{DOSand}(a) the gap has disappeared and the DOS has flattened.
Eventually the gap also closes for selective disorder by increasing the disorder strength, as already seen for $W=6t$ (green) in 
Fig.~\ref{DOSand}(b). This ``late'' gap closing correlates with a much robust topological phase, as we explain below.

The behavior of vacancy disorder 
equally distributed  among the two 
sublattices is similar to that of the Anderson disorder.  
The evolution of the Chern number as a function of
linear lattice size $d$ for various densities of vacancies $nv$ is shown in Fig.~\ref{Chern}(c). 
Notice the units in the vertical axis.  
The quantization disappears for a critical density $nv_c$ of 
around 4\%--6\%  compatible with the standard mechanism of ``levitation and 
pair annihilation'' of the extended states carrying the
Chern number. \cite{OAN07} A similar behavior has been reported in Ref.~\onlinecite{LHM14} for the 
case of vacancies. This mechanism is also supported by the analysis of the DOS
depicted in  Fig.~\ref{DOSvac}(a) 
%. The left hand side shows the DOS of the Haldane model 
for various density of vacancies randomly but equally distributed in the two sublattices. 
It is apparent that vacancies induce a finite
spectral weight inside the gap. As the disorder density increases
the impurity band widens and the center of the conduction and valence band that host the
extended states carrying the Chern number (1 and -1 respectively) move closer.  At a 
density of vacancies around $nv \approx 3\%$ the gap closes (inset of Fig.~\ref{DOSvac}(a)).
The Chern number remains quantized up to a density $nv_c$ ($4\% < nv_c < 6\%$)
where the ``annihilation'' occurs. Similar behavior has been
found in Ref.~\onlinecite{LHM14} although the critical density reported there is a bit different.


We will now discuss the results obtained for disorder selectively located in one of the  sublattices.
A comparison of the behavior of the Chern number as a function
of Anderson disorder strength for the two types of disorder, selective and equally distributed 
in the sublattices, is shown in Fig.~\ref{Chern}(b). The transition described above 
in the evenly distributed case (red squares) occurs sharply around $W_c\approx 5t$.
For this value, the Chern number in the selective disorder case (blue circles)
is still quantized to one. 
%Within numerical accuracy, the Chern number is well converged for $d\ge30$.
The topological transition occurs for disorder 
strengths one order of magnitude bigger (inset of Figure)
($40 t<W_c<50 t$) than in the equally distributed case. 

Similarly, for the case of disorder in the form of vacancies selectively distributed 
shown in Fig.~\ref{Chern}(d), the Chern number is still converging to one for
a density of vacancies of around 30\%. This is to be compared with 
the equally distributed disorder in the two sublattices (Fig.~\ref{Chern}(c)) 
where the graphic departs from 1 already at a critical density around 4\%--6\%.
%The case of the Anderson disorder (right hand side) is even more dramatic. 
The DOS for selective vacancy disorder, shown in Fig.~\ref{DOSvac}(b),
is also very different than in the previous case.  
Valence and conduction peaks do not move substantially as disorder  increases and the behavior of the gap
between the valence (conduction) and midgap bands is also different. As can be seen 
in the inset of Fig.~\ref{DOSvac}(b) the gap decreases and reopens well before the Chern number
ceases to be quantized what happens at a critical density of vacancies of $nv_c > 30\%$ 
(Fig.~\ref{Chern}(d)). At this density there is a noticeable gap 
separating the midgap band from the conduction and valence bands.

The case of disorder distributed in only one sublattice requires an alternative explanation.
We believe that both the topological transition and the localization mechanism 
are different in this case. The
robustness to disorder of the extended states carrying the Chern number is related to
the fact that one of the sublattices remains perfect. 
The component of the wave function corresponding to
the ``clean'' sublattice stays delocalized in a way
similar to the vacancy states of bilayer graphene 
described in Ref.~\onlinecite{CLV10}. 
The fact that the topological transition in the vacancies case occurs
for $nv_c > 30\%$,
%with the percolation transition of the honeycomb lattice \cite{Sta85} (around $30 \% $)
while the percolation transition of the triangular lattice takes place at 50\%, \cite{Sta85,SZ99}
points to the possibility that the topological transition occurs when the
disordered sublattice is destroyed and the extended (Bloch) state is driven 
by the underlying triangular lattice that is topologically trivial.

The case of vacancies is very intuitive: In the limit when we have removed
all the atoms in one of the triangular sublattices what remains is the 
extended wave function corresponding to the triangular sublattice.
Let us emphasize that we should not confuse the triangular lattice
with homogeneous complex hoppings, which break time reversal symmetry,
with the familiar triangular (or hexagonal) lattice with real hoppings.
A discussion of the transition between the honeycomb and two independent
triangular lattices is done in Appendix~\ref{gracieta}.
The corresponding bands are shown in Figs.~\ref{fig:spectrumTriang} 
and~\ref{fig:spectrumTriangReC} of the Appendix~\ref{gracieta}.

\section{Discussion and open issues}
%{\it Discussion and open issues.}
It is by now clear that topological features can change the localization behavior of the disordered electronic lattice systems. The topological transition (topological index going to zero or to a non-quantized value) and the localization transitions are related but do not occur, in general, at the same critical value of the disorder. In the ``ten-fold'' way classification of topological insulators, \cite{SRFL08} chiral classes (with well defined sublattice symmetry) do not support topological features in 2D and the issue of selective disorder (affecting only one sublattice) has only started to be explored. \cite{OPK14} The topological transition in the  2D (non chiral) classes  is assumed to occur through ``levitation and annihilation''  of the extended states carrying the topological index.\cite{OAN07} The mechanism is less understood in the chiral (sublattice symmetry) classes. \cite{RMLF12} In the one dimensional chiral case analyzed recently in Ref.~\onlinecite{PMHSP14}, the topological index $C$ remains quantized and non-fluctuating even when the bulk energy-spectrum is completely localized and after the insulating gap has closed. $C$ changes abruptly to a trivial value when disorder is further increased.

We observe a similar behavior for the Chern number $C$ for selective vacancy disorder:
$C$ remains quantized up to a large disorder strength but in our case extended states 
remain present possibly all the way until one of the sublattices is totally destroyed. 
This type of disorder realizes the transition from honeycomb lattice physics with topologically nontrivial  
features to the two topologically trivial uncoupled triangular lattices. 
%Mention the "Topological Anderson Insulators" \cite{LCJS09}.
Our results for the selective disorder do not convey with the standard
description of the topological transition based on ``levitation and annihilation''
of the extended states carrying the topological index.

%For a model system consisting of a tight-binding lattice Hamiltonian with ``topological bands" having nonzero Chern number, to which on-site disorder is added, we expect always a final regime for strong disorder where all states  are localized and have Chern number zero.

Sublattice symmetry plays a very subtle role in 2D topological insulators. 
It is a necessary ingredient in a minimal model to support non trivial topology
(a single band is necessarily trivial) but at the same time non trivial topology
requires it to be broken by NNN hoppings or otherwise.  
Ultimately, it is this broken chiral symmetry that is 
responsible for the sensitivity to selective disorder.

We understand  the topological transition as a transition from the honeycomb 
(topologically non--trivial for finite Haldane mass) to the two uncoupled triangular 
(trivial) sublattices within the same symmetry class. The ``localization transition'' 
is more complicated but in the limiting cases it is also between a 
Chern insulator (topologically non trivial) to a metal, again without changing 
symmetry or dimensionality.
The localization transition in graphene with vacancies equally distributed among the
two sublattices has been explored in some detail as a special 
example of chiral class \cite{OTetal10,HSetal14,FUH14} presenting some differences with the Anderson disorder. 
But the sensitivity of the 
localization transition to the sublattice selection 
has only been addressed more recently. \cite{LHM14,OPK14}
Reference~\onlinecite{LHM14} discusses the topological transition in the case of the time reversal symplectic class (Kane-Mele model). Our findings are qualitatively similar in the  equally diluted case but differ in the selected dilution case.  Ref.~\onlinecite{OPK14} analyses the localization properties of a 2D chiral systems (in particular in graphene) without topological features with emphasis on the selected dilution case.  Our results can not be directly compared with theirs because the NNN complex hoppings make a big difference. Apart from being responsible for the topological non trivial features, in the case of selective vacancy disorder, NNN provide the connectivity of the  unaffected triangular sublattice. In the extreme case  when one sublattice is totally depleted, the remaining sublattice in our case is metallic with bands shown in Fig. \ref{fig:spectrumTriangReC} of Appendix \ref{gracieta}. It will be interesting to analyze the scaling behavior of the Haldane model along the lines of Ref. \onlinecite{OPK14}. A recent calculation of the conductivity tensor in the disordered Haldane model suggests that an asymmetry between sublattices
A and B can help to stabilize the Chern insulator, \cite{GCR15} an intriguing result compatible with our findings.
It is interesting to note that the Haldane model  with purely imaginary hoppings ($\phi=\pi/2$) has particle--hole symmetry so that it belongs to class D whose Chern number also belongs  to ${\mathbb Z}$.\footnote {We thank Fernando de Juan for this observation.} We do not expect significant differences in the results for  the case of a general value of $\phi$.

\begin{acknowledgments}
We gratefully acknowledge useful conversations with Bel\'en Valenzuela on the transition from the honeycomb to the Hexagonal lattice. 
We also thank Fernando de Juan for a critical reading of the manuscript and useful suggestions.
EC acknowledges the financial support of FCT-Portugal through grant No. EXPL/FIS-NAN/1720/2013. This research was supported in part by the Spanish MECD grants FIS2011-23713, PIB2010BZ-00512, the  European Union structural funds and the Comunidad de Madrid MAD2D-CM Program (S2013/MIT-3007),   the European Union Seventh Framework Programme under grant agreement no. 604391 Graphene Flagship FPA2012-32828agship.
\end{acknowledgments}
%\bibliography{Chern}

\appendix

\section{From honeycomb to two triangular lattices}
\label{gracieta}
\begin{figure*}
\begin{centering}
\includegraphics[width=1.6\columnwidth]{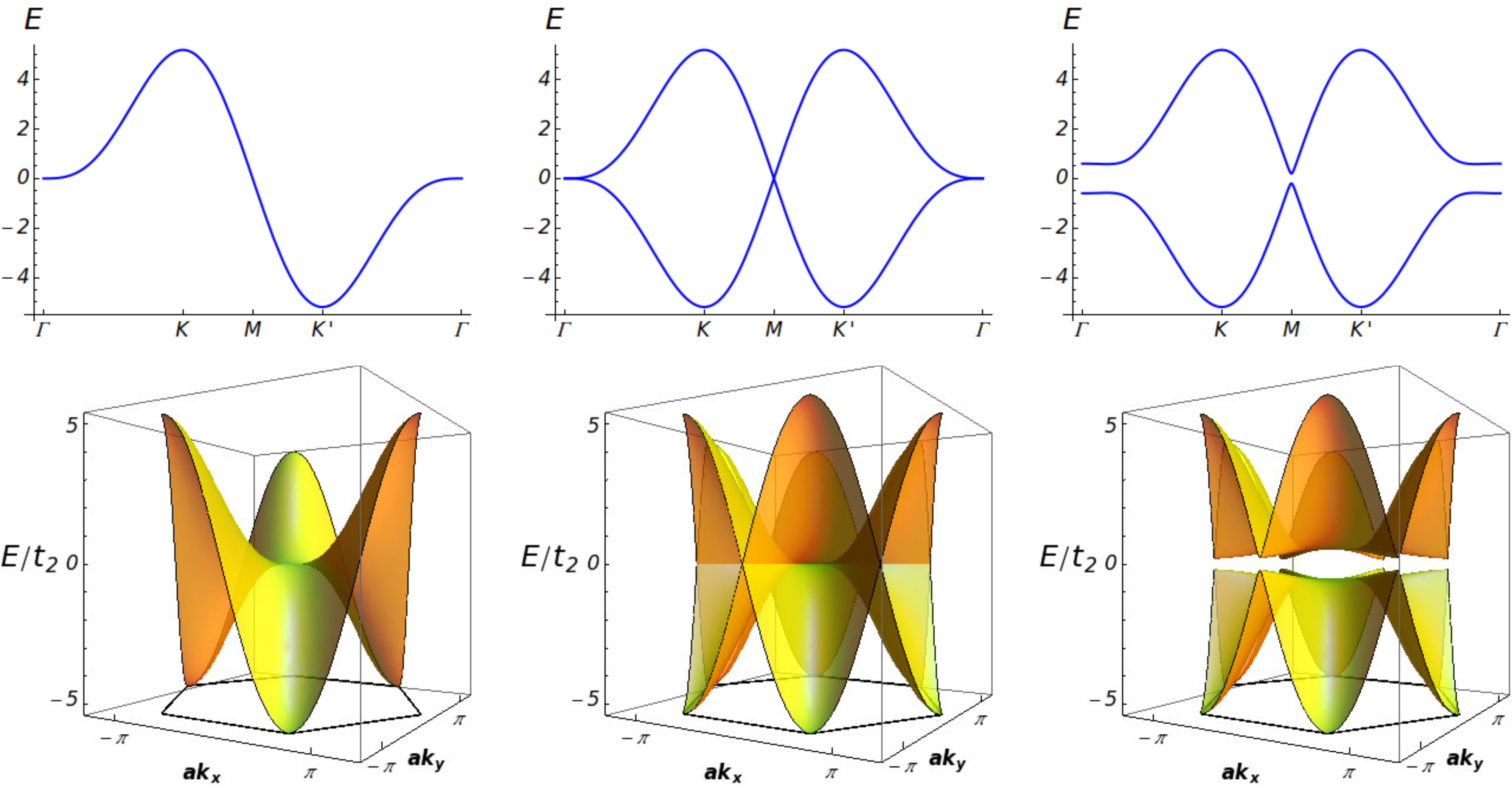}
%\includegraphics[width=0.6\columnwidth]{Fig4a}~~
%\includegraphics[width=0.6\columnwidth]{Fig4b}~~
%\includegraphics[width=0.6\columnwidth]{Fig4c}\\
%\vspace{0.5cm}
%\includegraphics[width=0.65\columnwidth]{Fig4d}~~~
%\includegraphics[width=0.65\columnwidth]{Fig4e}~~~
%\includegraphics[width=0.65\columnwidth]{Fig4f}
\par\end{centering}
\caption{\label{fig:spectrumTriang}Energy spectrum for the triangular lattice
with imaginary hoppings (left), two decoupled triangular lattices
with imaginary hoppings related by complex conjugation (center), and
coupled triangular lattices (right) as given by Eq.~\eqref{eq:EkHaldaneIm}.
Top panels are the same as bottom panels along a particular path in
the BZ.}
\end{figure*}
%
%\begin{figure*}
%\begin{centering}
%\includegraphics[width=0.45\columnwidth]{Fig5a}~~~~~~~~~~~~~~~~~\includegraphics[width=0.45\columnwidth]{Fig5b}~
%\includegraphics[width=0.5\columnwidth]{Fig5c}
%\par\end{centering}
%
%\caption{\label{fig:spectrumTriangReC}(left) Chern number for the Haldane
%model as a function of the parameter $t/t_{2}$. ( middle, right)
%Energy spectrum for the triangular lattice with real hoppings. }
%\end{figure*}
%

\begin{figure*}
\begin{centering}
\includegraphics[width=1.6\columnwidth]{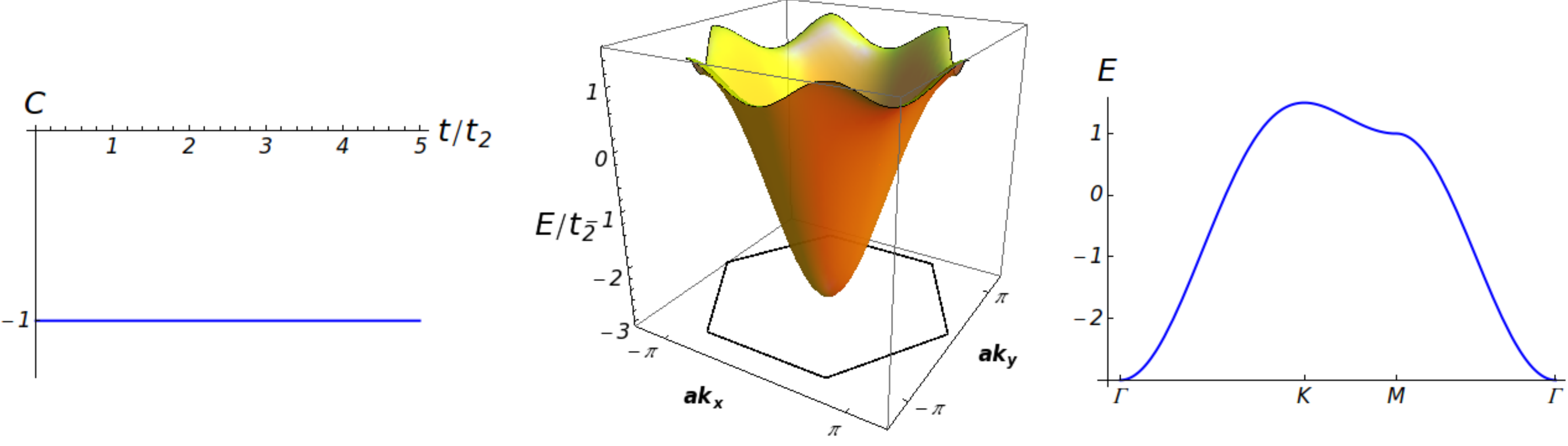}
\par\end{centering}
\caption{\label{fig:spectrumTriangReC}(left) Chern number for the Haldane
model as a function of the parameter $t/t_{2}$. (middle)
Energy spectrum for the triangular lattice with real hoppings.
(right) The same as the middle panel along a particular path in
the BZ. }
\end{figure*}

Dilution of a single sublattice leads, after 100\% dilution, to the
triangular lattice. Since the latter has a single band, the Chern
number must be zero. In this section we analyze whether the transition
found for the selected dilution case can be attributed to the fact
that, for some critical density of vacancies, the system is already
more triangular lattice like and therefore the Chern number is zero.

To address this possibility we study how the gap and Chern number
evolves with the NN hopping $t$. For $t=0$ we end up with the triangular
lattice. The spectrum and the eigenvectors are obtained by the $2\times2$
Hamiltonian given by

\[
\mathcal{H}_{\mathbf{k}}=\left(\begin{array}{cc}
m(\mathbf{k}) & f(\mathbf{k})\\
f^{*}(\mathbf{k}) & -m(\mathbf{k})
\end{array}\right),
\]
where
\begin{eqnarray*}
m(\mathbf{k}) & = & -2t_{2}[\sin(\mathbf{k}\cdot\mathbf{a}_{1})-\sin(\mathbf{k}\cdot(\mathbf{a}_{1}-\mathbf{a}_{2}))-\sin(\mathbf{k}\cdot\mathbf{a}_{2})],\\
f(\mathbf{k}) & = & t(1+e^{i\mathbf{k}\cdot\mathbf{a}_{1}}+e^{i\mathbf{k}\cdot\mathbf{a}_{2}})\,,
\end{eqnarray*}
and we have chosen $\mathbf{a}_{1} = a(-1,\sqrt{3})/2$ and $\mathbf{a}_{2} = a(1,\sqrt{3})/2$
 with $a$ the lattice constant for the underlaying hexagonal lattice.
The spectrum is easily obtained and reads
\begin{equation}
E_{\mathbf{k}}=\pm\sqrt{m(\mathbf{k})^{2}+|f(\mathbf{k})|^{2}}\,.\label{eq:EkHaldaneIm}
\end{equation}
In the limit $t/t_{2}\rightarrow0$ we get two decoupled triangular
lattices with imaginary hoppings related by complex conjugation, the
spectrum being $E_{\mathbf{k}}=\pm m(\mathbf{k})$. In Fig.~\ref{fig:spectrumTriang}
we show the spectrum $E_{\mathbf{k}}=m(\mathbf{k})$ for the triangular
lattice with imaginary hoppings (left), the spectrum $E_{\mathbf{k}}=\pm m(\mathbf{k})$
for two decoupled triangular lattices with imaginary hoppings (center),
and the spectrum given by Eq.~\eqref{eq:EkHaldaneIm} when the system
is perturbed by a finite $t$ leading to a finite gap (right).

It is apparent that any finite NN hopping parameter $t$ lifts the
degeneracy that can be seen in the middle panels of Fig.~\ref{fig:spectrumTriang},
and opens a finite gap in the spectrum. Since we know by inspection
that the gap remains open for any $t/t_{2}$, including $t/t_{2}\gg1$
where the system in topologically non-trivial, then we conclude that
it must be non-trivial for any finite $t$. This is explicitly shown
in the left panel of Fig.~\ref{fig:spectrumTriangReC}, where we
represent the Chern number for the Haldane model as a function of
$t/t_{2}$. The Chern number has been obtained using Fukui's method.\cite{FHS05} 
So, the system is trivial only for $t=0$, when the
two triangular lattices are decoupled.

Let us emphasize that we should not confuse the triangular lattice
with homogeneous complex hoppings, which breaks time reversal symmetry,
with the familiar triangular (or hexagonal) lattice with real hoppings.
The spectrum for the latter reads
\begin{eqnarray*}
E_{\mathbf{k}} & = & -t_{2}\left(\cos(\mathbf{k}\cdot\mathbf{a}_{2})+\cos(\mathbf{k}\cdot\mathbf{a}_{1})+\cos[\mathbf{k}\cdot(\mathbf{a}_{1}-\mathbf{a}_{2})]\right)\\
 & = & -t_{2}\left(2\cos\left(\frac{ak_{x}}{2}\right)\cos\left(\frac{\sqrt{3}}{2}ak_{y}\right)+\cos(ak_{x})\right)\,.
\end{eqnarray*}
Such energy spectrum is shown in Fig.~\ref{fig:spectrumTriangReC}
(middle and right panels).

%\end{widetext}
\bibliography{Chern}
\end{document}